\documentclass[]{aastex631}

\usepackage{booktabs}

\newcommand{\sw}[1]{\texttt{#1}}

\begin{document}

\title{A Near-IR Search for Helium in the Superluminous Supernova SN\,2024ahr}

\author[0000-0003-0871-4641]{Harsh Kumar}
\affiliation{Center for Astrophysics \textbar{} Harvard \& Smithsonian, 60 Garden Street, Cambridge, MA 02138-1516, USA}
\affiliation{The NSF AI Institute for Artificial Intelligence and Fundamental Interactions, USA}

\author[0000-0002-9392-9681]{Edo Berger}
\affiliation{Center for Astrophysics \textbar{} Harvard \& Smithsonian, 60 Garden Street, Cambridge, MA 02138-1516, USA}
\affiliation{The NSF AI Institute for Artificial Intelligence and Fundamental Interactions, USA}

\author[0000-0003-0526-2248]{Peter K.~Blanchard}
\affiliation{Center for Astrophysics \textbar{} Harvard \& Smithsonian, 60 Garden Street, Cambridge, MA 02138-1516, USA}
\affiliation{The NSF AI Institute for Artificial Intelligence and Fundamental Interactions, USA}

\author[0000-0001-6395-6702]{Sebastian Gomez}
\affiliation{Center for Astrophysics \textbar{} Harvard \& Smithsonian, 60 Garden Street, Cambridge, MA 02138-1516, USA}

\author[0000-0002-1125-9187]{Daichi Hiramatsu}
\affiliation{Center for Astrophysics \textbar{} Harvard \& Smithsonian, 60 Garden Street, Cambridge, MA 02138-1516, USA}
\affiliation{The NSF AI Institute for Artificial Intelligence and Fundamental Interactions, USA}

\author{Moira Andrews} 
\affiliation{Las Cumbres Observatory, 6740 Cortona Drive, Suite 102, Goleta, CA 93117-5575, USA}
\author{K. Azalee Bostroem}  
\affiliation{Las Cumbres Observatory, 6740 Cortona Drive, Suite 102, Goleta, CA 93117-5575, USA} 
\author[0000-0002-7937-6371]{Yize Dong} 
\affiliation{Center for Astrophysics \textbar{} Harvard \& Smithsonian, 60 Garden Street, Cambridge, MA 02138-1516, USA}
\affiliation{Department of Physics and Astronomy, University of California, 1 Shields Avenue, Davis, CA 95616-5270, USA}

\author{Joseph Farah}   
\affiliation{Las Cumbres Observatory, 6740 Cortona Drive, Suite 102, Goleta, CA 93117-5575, USA} 
\affiliation{Department of Physics, University of California, Santa Barbara, CA 93106-9530, USA}
\author{Estefania Padilla Gonzalez}  
\affiliation{Las Cumbres Observatory, 6740 Cortona Drive, Suite 102, Goleta, CA 93117-5575, USA}
\affiliation{Department of Physics, University of California, Santa Barbara, CA 93106-9530, USA}
\author{D. Andrew Howell}   
\affiliation{Las Cumbres Observatory, 6740 Cortona Drive, Suite 102, Goleta, CA 93117-5575, USA} 
\affiliation{Department of Physics, University of California, Santa Barbara, CA 93106-9530, USA}
\author{Curtis McCully} 
\affiliation{Las Cumbres Observatory, 6740 Cortona Drive, Suite 102, Goleta, CA 93117-5575, USA} 
\author[0009-0008-9693-4348]{Darshana Mehta} 
\affiliation{Department of Physics and Astronomy, University of California, 1 Shields Avenue, Davis, CA 95616-5270, USA}

\author{Megan Newsome}
\affiliation{Las Cumbres Observatory, 6740 Cortona Drive, Suite 102, Goleta, CA 93117-5575, USA} 
\affiliation{Department of Physics, University of California, Santa Barbara, CA 93106-9530, USA}

\author[0000-0002-7352-7845]{Aravind P. Ravi}
\affiliation{Department of Physics and Astronomy, University of California, 1 Shields Avenue, Davis, CA 95616-5270, USA}

\author{Giacomo Terreran} 
\affiliation{Las Cumbres Observatory, 6740 Cortona Drive, Suite 102, Goleta, CA 93117-5575, USA}

\begin{abstract}

We present a detailed study of SN\,2024ahr, a hydrogen-poor superluminous supernova (SLSN-I), for which we determine a redshift of $z=0.0861$.  SN\,2024ahr has a peak absolute magnitude of $M_g\approx M_r\approx -21$ mag, rest-frame rise and decline times (50\% of peak) of about 40 and 80 days, respectively, and typical spectroscopic evolution in the optical band. Similarly, modeling of the UV/optical light curves with a magnetar spin-down engine leads to typical parameters: an initial spin period of $\approx 3.3$ ms, a magnetic field strength of $\approx 6\times 10^{13}$ G, and an ejecta mass of $\approx 9.5$ M$_\odot$. Due to its relatively low redshift we obtained a high signal-to-noise ratio near-IR spectrum about 43 rest-frame days post-peak to search for the presence of helium.  We do not detect any significant feature at the location of the \ion{He}{1}$\,\lambda 2.058$ $\mu$m feature, and place a conservative upper limit of $\sim 0.05$ M$_\odot$ on the mass of helium in the outer ejecta.  We detect broad features of \ion{Mg}{1}$\,\lambda 1.575$ $\mu$m and a blend of \ion{Co}{2}$\,\lambda 2.126$ $\mu$m and \ion{Mg}{2}$\,\lambda 2.136$ $\mu$m, which are typical of Type Ic SNe, but with higher velocities.  Examining the sample of SLSNe-I with NIR spectroscopy, we find that, unlike SN\,2024ahr, these events are generally peculiar. This highlights the need for a large sample of prototypical SLSNe-I with NIR spectroscopy to constrain the fraction of progenitors with helium (Ib-like) and without helium (Ic-like) at the time of explosion, and hence the evolutionary path(s) leading to the rare outcome of SLSNe-I.
\end{abstract}


\section{Introduction} 
\label{sec:intro}

Modern wide-field optical time-domain surveys have enabled the discovery of a rare class of transients dubbed ``hydrogen poor superluminous supernovae" (SLSNe-I), which can exceed the luminosities of normal stripped-envelope core-collapse Supernovae (Type Ib and Ic SNe) by two orders of magnitude \citep{2007ApJ...668L..99Q, 2009Natur.462..624G, 2009ApJ...690.1358B, 2011Natur.474..487Q, 2011ApJ...743..114C, 2012Sci...337..927G, 2016ApJ...820L..38S, 2019ARA&A..57..305G, 2021A&G....62.5.34N, 2024MNRAS.535..471G}. SLSNe-I also exhibit systematically longer timescales and bluer colors at an early time and distinct optical/UV spectra near peak and in the nebular phase compared to normal SNe Ib/c \citep{2016MNRAS.458.3455M, 2018ApJ...854...37S, 2019ApJ...882..102G}.  Taken together, these properties point to a different power source in SLSNe-I, namely a highly magnetized, rapidly spinning neutron star (magnetar; \citealt{2010ApJ...717..245K, 2010ApJ...719L.204W}), as opposed to radioactive decay of $^{56}$Ni in normal SNe Ib/c.  In addition, SLSNe-I are rare ($\lesssim 0.3\%$ of SESNe (type Ib/c) \citealt{2013MNRAS.431..912Q, 2021MNRAS.500.5142F}), exhibit a distinct pre-explosion mass distribution \citep{2020ApJ...897..114B}, and occur in different host galaxy and local environments than SNe Ib/c \citep{2015MNRAS.452.1567C, 2016ApJ...830...13P}, pointing to a restricted evolutionary path to their formation.

Although there are substantial differences between SLSNe-I and SNe Ib/c, both types of explosions arise from stripped massive stars, and it is therefore natural to explore whether SLSNe-I also exhibits a sub-class that is Ib-like (retaining a helium layer) or if these events are only Ic-like (stripped of helium). The presence of helium can in principle be probed in the optical regime, using \ion{He}{1} lines at 3888, 5875, 6678, and  7065 \AA\ \citep{2001AJ....121.1648M, 2002AAS...200.1404G, 2020ApJ...902L...8Y}, but these lines tend to be blended with other, stronger features (e.g., \ion{Ca}{2}$\,\lambda 3934,3968$\AA, \ion{Na}{1}$\,\lambda 5875-5890$\AA, and \ion{C}{2}$\,\lambda 6580,7234$\AA), especially in the case of SLSNe-I, which have large ejecta velocities and hence broad features.  Additionally, the optical lines are expected to be weak and may thus be present only for large helium shell masses ($\gtrsim 0.2 - 1.0$ M$_\odot$; \citealt{2020MNRAS.499..730T}).  Taken together, this limits claims of helium detections in the optical band from being robust \citep{2020ApJ...902L...8Y}.

On the other hand, as shown in the case of SNe Ib/c, NIR spectroscopy provides a more robust probe for the presence of helium \citep{2023A&A...675A..83H}. In particular, the two strongest \ion{He}{1} lines in the NIR are at 1.083 and 2.058 $\mu$m, with a third weaker line at 1.700 $\mu$m. However, as in the optical band, the 1.083 $\mu$m line suffers from blending with \ion{C}{1} and \ion{Mg}{2}, and it is therefore present in both SNe Ib and Ic, even if the latter lack helium \citep{2022ApJ...925..175S}.  The 2.058 $\mu$m line, however, is free of strong blending and thus provides the most robust signature for helium (e.g., \citealt{2001AJ....121.1648M, 2021ApJ...908..150W,2022ApJ...925..175S}).  Indeed, models of SNe Ib/c spectra have shown that even a nearly bare CO core, with a helium mass down to $\sim 0.1$ M$_\odot$, can still lead to a detectable signature at 2.058 $\mu$m \citep{2012MNRAS.422...70H, 2020MNRAS.499..730T}.

Only a few attempts have been made to obtain NIR spectra of SLSNe-I covering the \ion{He}{1}$\,\lambda 2.058$ $\mu$m line, predominantly in the nearest events (SN\,2017egm: \citealt{2023ApJ...949...23Z}; SN\,2018bsz: \citealt{2022A&A...666A..30P}; SN\,2018ibb: \citealt{2024A&A...683A.223S}; and SN\,2019hge: \citealt{2023ApJ...943...41C}. SN\,2019hge).  These NIR searches resulted predominantly in non-detection, with the single exception of a claimed detection in SN\,2019hge at $z=0.0866$ \citep{2020ApJ...902L...8Y}; we note that this event is one of the lowest luminosity SLSNe-I detected to date, comparable to the luminous end of normal SESNe \citep{2022ApJ...941..107G}.

Here, we present a detailed study of the relatively nearby ($z\approx 0.08$) SLSN-I event SN\,2024ahr, including NIR spectroscopy to search for the presence of helium.  We demonstrate that SN\,2024ahr is typical of the luminous SLSN-I population, and show that its NIR spectrum is similar to a SN Ic (albeit with broader features). 
The paper is arranged as follows. In \S\ref{sec:obs} we present the discovery of SN\,2024ahr and our UV, optical, and NIR follow-up observations.  In \S\ref{sec:analysis} we explore the photometric properties and model the light curves with a magnetar central engine model; discuss the spectroscopic evolution; and analyze the NIR spectrum, including a comparison to the small sample of existing NIR spectra of SLSNe-I, and determination of a limit on the helium mass.  We summarize the key results in \S\ref{sec:conclusion}.

\section{Discovery and Observations}
\label{sec:obs}

\subsection{Discovery and Classification} 
\label{sec:discovery}

\begin{figure}[t!]
\center
\includegraphics[width=0.9\linewidth]{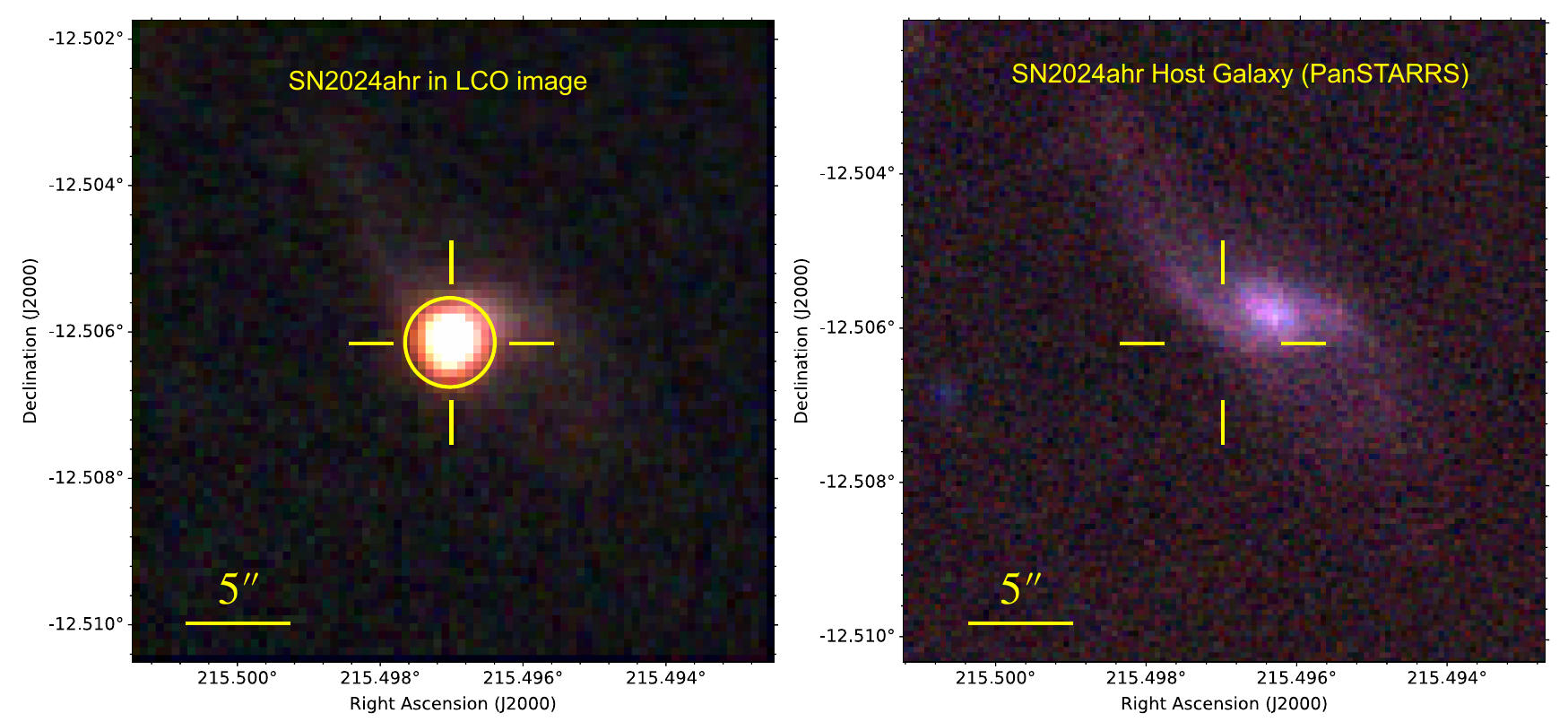}
\caption{{\it Left:} RGB image of SN\,2024ahr from our Las Cumbres Observatory data, about 59 days after discovery (yellow circle). {\it Right:} Pre-discovery image of the host galaxy from PanSTARRS-DR2 data \citep{2018AAS...23143601F}. SN\,2024ahr is located $\approx 2.45^{\prime \prime}$ ($\approx 4.1$ kpc) from the center of the host galaxy near a spiral arm.}
\label{fig:discovery}
\end{figure}

SN\,2024ahr was discovered by the Zwicky Transient Facility~\citep[ZTF;][]{2014htu..conf...27B, 2019PASP..131a8002B} on 2024 January 16 (MJD = 60325), with an internal designation ZTF24aacrbua. It was also detected by ATLAS (ATLAS24bwz), Pan-STARRS (PS24aun), Gaia (Gaia24apk), and BlackGem (BGEM J142159.27$-$123021.9) independently\footnote{\url{https://www.wis-tns.org/object/2024ahr/}}. The discovery magnitude was $m_g = 20.98 \pm 0.33$, with a position of R.A.=$00^{h} 24^{m} 34.71^{s}$, Decl.=$+47^{\circ} 13^{\prime} 21.^{\prime\prime}55$ (J2000) \citep{2024TNSTR.168....1F}. Images of SN\,2024ahr from our follow-up observations and its underlying host galaxy from archival Pan-STARRS DR2 are shown in Figure~\ref{fig:discovery}. 

\citet{2024TNSCR.619....1W} obtained a spectrum of SN\,2024ahr on 2024 March 6 (50 days post-discovery), and classified it as a SLSN-I at an approximate redshift of $z\approx 0.1$, based on template matches to previous SLSNe-I. Here we refine the redshift to $z=0.08612 \pm 0.00004$ based on host-galaxy emission lines.

\subsection{Optical Imaging Observations}

\paragraph{Las Cumbres Observatory (LCO)}

We conducted follow-up imaging observations with the Sinistro cameras on the 1-meter telescopes in the Las Cumbres Observatory (LCO; \citealt{2013PASP..125.1031B}) network starting on 2024 March 8 to September 15 (192 days) The observations were part of the Global Supernova Project \citep{2017AAS...23031803H} and utilized the $B,V,g,r,i$ filters. Photometric measurements were made using the point-spread function (PSF) fitting, employing the \texttt{lcogtsnpipe} pipeline \citep{2016MNRAS.459.3939V}. The $g,r, i$ images were processed with the ZOGY image subtraction algorithm \citep{2016ApJ...830...27Z} using PanSTARRS DR2 image as templates. In the absence of reference templates in $B$ and $V$ bands, we report the PSF photometry. Given the blue color of the galaxy and the brightness of the source, we do not expect the lack of image subtraction to affect the photometry significantly in these bands.

\paragraph{Zwicky Transient Facility (ZTF)} The ZTF observations cover the time range from discovery on 2024 January 16 to July 28 (195 days) in the $g,r$ bands. We obtained the ZTF photometry from the Automatic Learning for the Rapid Classification of Events (ALeRCE) broker \citep{2021AJ....161..242F}.

\paragraph{Asteroid Terrestrial-impact Last Alert System (ATLAS)} The ATLAS observations cover the time range from 2024 January 16 to July 28 (195 days) in the $c$ band and from 2024 January 15 to August 2 (201 days) in the $o$-band. We obtained ATLAS photometry through the ATLAS Forced Photometry Server~\citep{2021TNSAN...7....1S}, selecting data with a signal-to-noise ratio of $\geq 5$. 

\begin{figure}[t!]
\center
\includegraphics[width=0.9\linewidth]{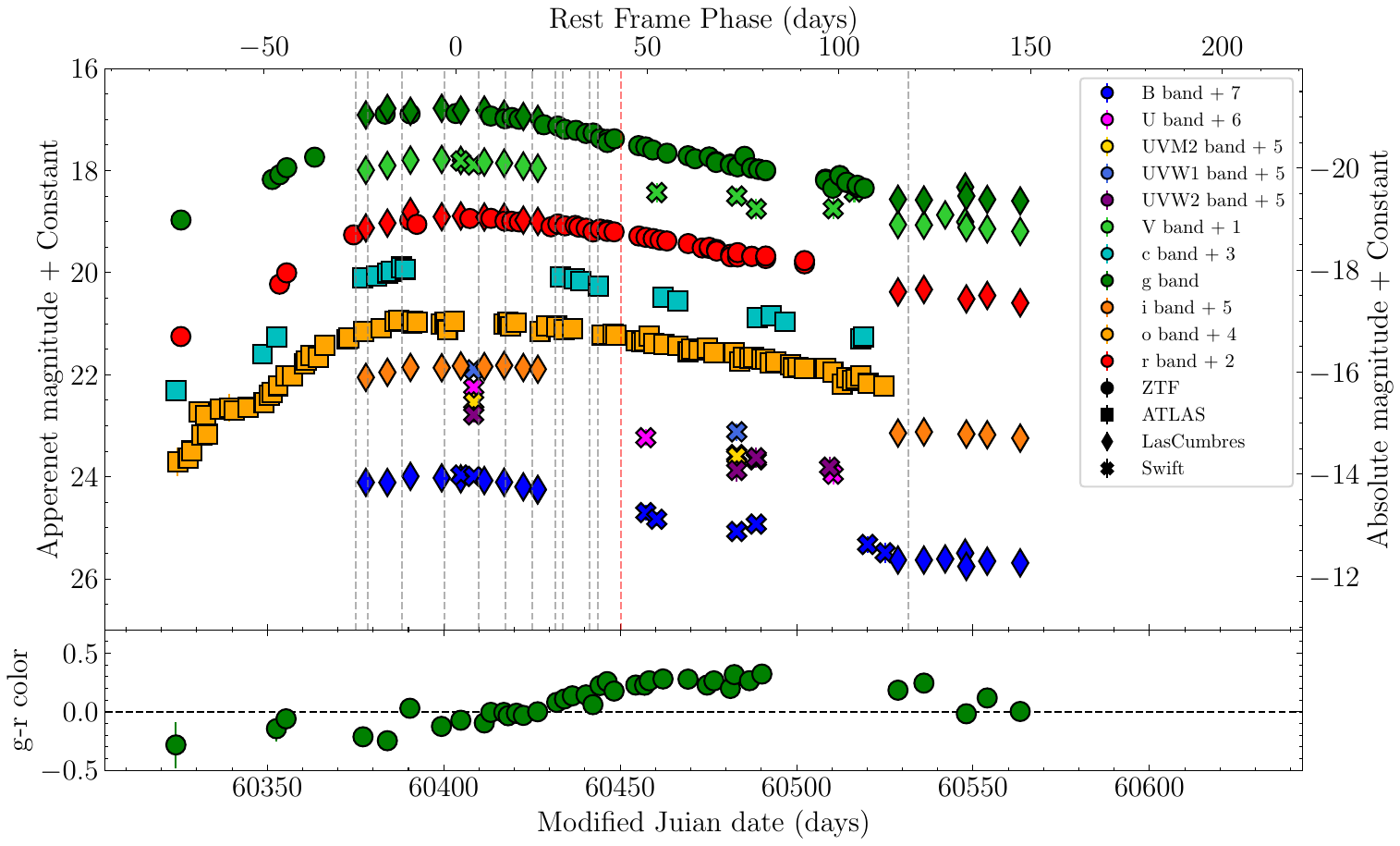}
\caption{Optical and UV light curves of SN\,2024ahr. All magnitudes are plotted in AB system and are corrected for Galactic Extinction. Vertical lines mark the epochs of optical (gray lines) and NIR (red line) spectroscopy. SN\,2024ahr rises from discovery to peak in $\approx 78$ d and reaches a peak absolute magnitude of $m_{g}\approx m_r\approx -21.0$. The $g-r$ color (bottom) gradually reddens from about $-0.3$ to $+0.3$ mag from the rise to the decline of the light curve. The $o$-band light curve shows an early bump.}
\label{fig:lc}
\end{figure}

\paragraph{Neil Gehrels {\it Swift} Observatory}
We obtained ultraviolet and optical observations using the UV/Optical Telescope (UVOT; \citealt{2005SSRv..120...95R}) in six filters (UVW2, UVM2, UVW1, $U$, $B$, $V$) in several epochs spanning from 2024 April 4 to August 3 (123 days). We performed aperture photometry using the \texttt{UVOTSOURCE} pipeline with an aperture of $5''$ radius, and included standard aperture corrections. We used a signal-to-noise ratio of $\geq 5$ as the detection limit in all bands. Since the transient was still bright when the field became Sun-constrained, no template images are available for image subtraction. Therefore, the photometry at later epochs (MJD $>$ 60450) in the UV filters is probably contaminated by host emission and is excluded from the subsequent analysis.

The light curves of SN\,2024ahr are shown in Figure~\ref{fig:lc}. Photometric measurements in the $g,c,r,i$, and $o$ bands are in AB magnitudes, and all other photometry is in the Vega system unless stated otherwise. All measurements have been corrected for Galactic extinction, with $E(B-V) = 0.092$ mag \citep{2011ApJ...737..103S}, assuming the~\cite{1999PASP..111...63F} reddening law with $R_V = 3.1$. Throughout this work, we use standard Planck18 FlatLambdaCDM cosmology~\citep{2020A&A...641A...6P} with Hubble constant $H_{0} = 67.4 \pm 0.5~\mathrm{km~s^{-1}~Mpc^{-1}}$; matter density parameter $\Omega_{m}= 0.315 \pm 0.007$. We estimated a luminosity distance of $d_{L} = 405\pm 19$ Mpc using a redshift $z = 0.08612\pm 0.00004$ (see \S~\ref{sec:spec}) for SN\,2024ahr.

\begin{figure}[t!]
\center\includegraphics[width=0.75\linewidth]{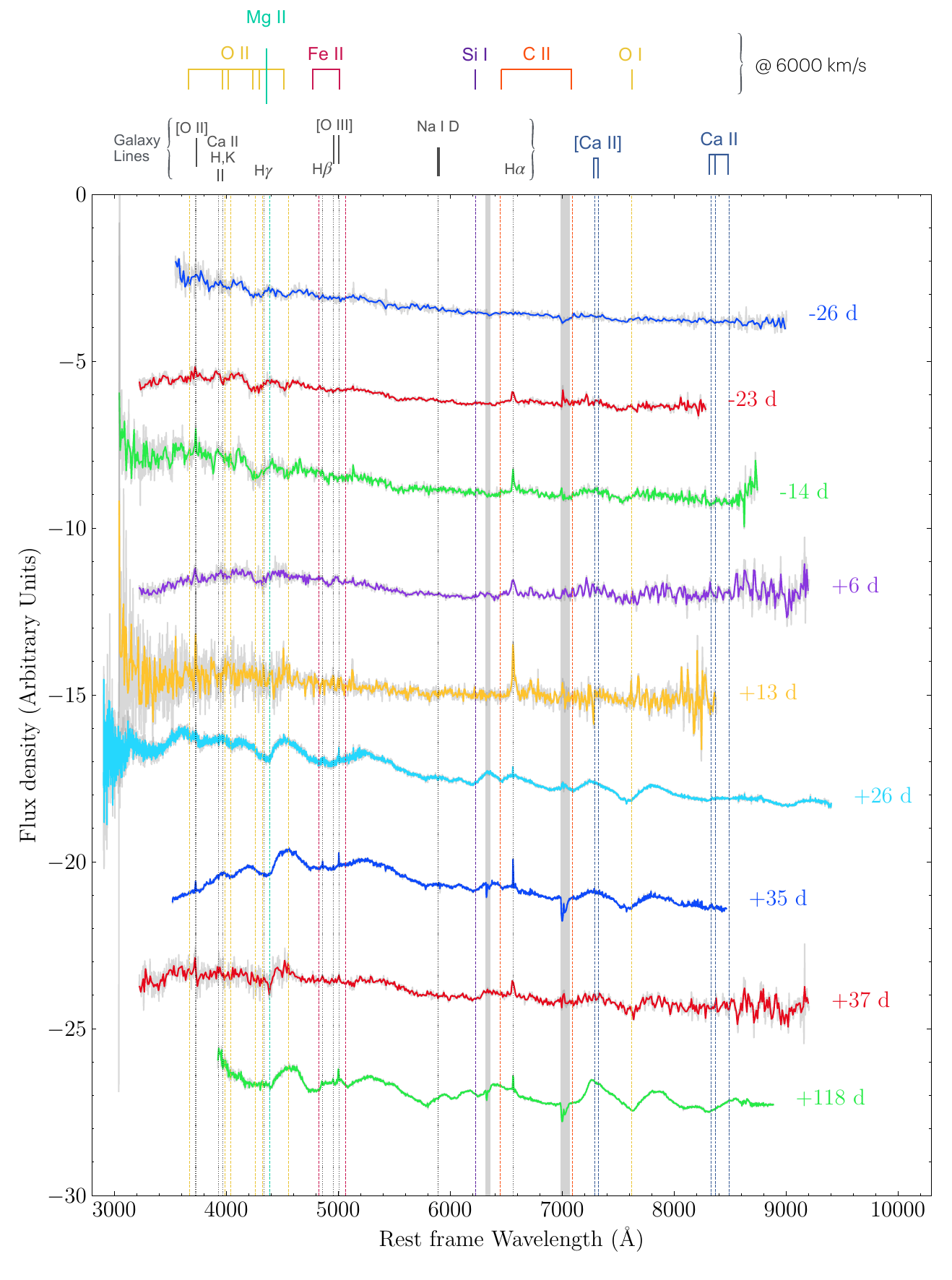}
\caption{Optical spectra of SN\,2024ahr, spanning phases of $-26$ to $+118$ d. The spectra exhibit a blue continuum at early time with \ion{O}{2} and \ion{Fe}{2}/\ion{Fe}{3} lines typical of SLSNe-I. At later times, the spectra are dominated by various ionization states of O, Fe, C, and Ca with a typical velocity of $\sim 6000$ km s$^{-1}$. The blueshifted position of features is marked with $v = 6000$ km s$^{-1}$ estimated using the spectrum at $+35$ d.}
\label{fig:speceval}
\end{figure}

\subsection{Optical Spectroscopy}
\label{sec:spec}
Following the SLSN-I classification of SN\,2024ahr, we obtained 6 epochs of LCO optical spectra with the FLOYDS spectrographs mounted on the 2m Faulkes Telescope North (FTN) and South (FTS) at Haleakala (USA) and Siding Spring (Australia), respectively, spanning from 2024 March 9 to May 14. The observations were made as a part of the Global Supernova Project \citep{Howell2017AAS...23031803H}. We used a $2\arcsec$ slit placed on the target along the parallactic angle \citep{Filippenko1982}. One-dimensional spectra were extracted, reduced, and calibrated following standard procedures using \texttt{floyds\_pipeline}\footnote{\url{https://github.com/LCOGT/floyds_pipeline}} \citep{Valenti2014}. The resulting spectra are plotted in Figure~\ref{fig:speceval}.

We also obtained a spectrum as a part of follow-up observations on 2024 May 1 with the Low-Resolution Imaging Spectrometer \citep[LRIS;][]{1995PASP..107..375O} mounted on the Keck~I 10-m telescope. The observations were performed with a 1$\arcsec$ wide slit, the 600/4000 grism, 400/8500 grating, and a 560 dichroic setup. The data were reduced using standard procedures in the fully-automated reduction pipeline for LRIS longslit spectra, \verb |LPipe| \citep{2019PASP..131h4503P}. The spectrum covers a wavelength range of $3824-9197$ \AA\ with a resolution of $\approx 1500$.

We obtained an additional spectrum with the Binospec spectrograph \citep{2019PASP..131g5004F} on the 6.5-m MMT telescope on 2024 May 11. We used a combination of 270 lines/mm grating with $1^{\prime \prime}$ wide slit and the LP3800 filter, covering a wavelength range of $3825-9200$ \AA\ with a resolution of $\approx 1500$. The data was reduced using \sw{PyRAF} following standard procedure via the MMT pipeline. The one-dimensional spectrum was extracted and flux-calibrated using standard star observation obtained at the same configuration close in time to the science observations.

Finally, we obtained a spectrum using the Low Dispersion Survey Spectrograph 3~\citep[LDSS3;][]{2016ApJ...817..141S} on the 6.5-m Magellan Clay telescope on 2024 August 9. We used the VPH-All grism with a 1$\arcsec$ wide slit, covering a wavelength range of $4264 - 9652$ \AA\ with a resolution of $\approx 700$. The spectrum was bias-subtracted and flat-fielded, and the sky background was modeled and subtracted from the 2D image. The one-dimensional spectra were extracted, and a wavelength calibration was applied using an arc lamp spectrum taken directly after the science image. Relative flux calibration was applied using standard star observation obtained at the same configuration close in time to the science observations.

The spectra (see Figure~\ref{fig:speceval}) exhibit multiple host galaxy emission lines (H$\alpha$, H$\beta$, H$\gamma$, [\ion{N}{2}], [\ion{O}{2}], [\ion{O}{3}] and [\ion{Na}{1} D]) from which we determine a redshift of $z=0.08612 \pm 0.00004$.

\subsection{Near-IR Spectroscopy}

\begin{figure}[t!]
\center\includegraphics[width=0.9\linewidth]{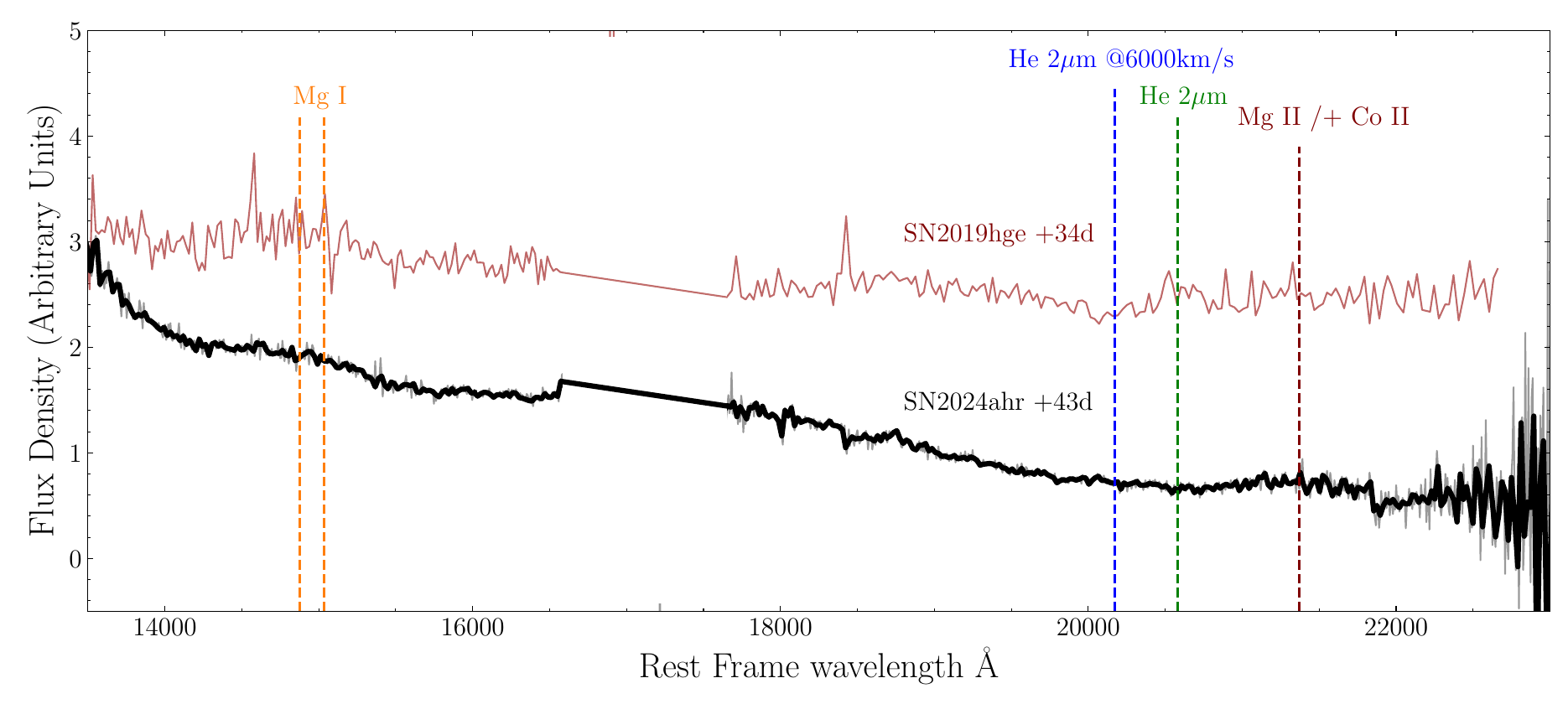}
\caption{NIR spectrum of SN\,2024ahr obtained at a phase of +43 d (black). The Spectrum exhibits broad features of \ion{Mg}{1} ($\sim 1.5$ $\mu$m) and \ion{Mg}{2} + \ion{Co}{2} ($\sim 2.13$ $\mu$m). We do not detect significant absorption from \ion{He}{1} at $2\mu$m. For comparison, we also show the NIR spectrum of SN\,2019hge (brown), which appears to exhibit the \ion{He}{1}$\,\lambda 2.058$ $\mu$m line with a blueshift of $\approx 6000$ km s$^{-1}$ \citep{2020ApJ...902L...8Y}.}
\label{fig:specnir}
\end{figure}

Given the relatively low redshift of SN\,2024ahr, we initiated a search for helium lines by obtaining a NIR spectrum with the FLAMINGOS-2 spectrograph on the Gemini-South 8-m telescope on 2024 May 20 through a Director’s Discretionary Time (DDT) program (GS-2024A-DD-106, PI: Kumar). We used the HK Filter and disperser pair with a focal plane mask of the 3-pixel slit (0.45) at low read-noise to obtain multiple science exposures totaling about 2 hours on the source. The slit was aligned to the parallactic angle to minimize atmospheric dispersion. The HK grating provides coverage at $1.45-2.45$ $\mu$m.  We reduced the data using the \sw{PypeIt} package~\citep{pypeit:joss_pub}. The spectral images were flat-fielded using GCALflats to correct pixel-to-pixel sensitivity variations. Wavelength calibration was achieved using OH sky emission lines that are well-distributed across the NIR range. A flux calibration was applied using a sensitivity function estimated from two nearby A0V standard stars. The resulting 1D spectra were co-added, and the spectrum was corrected for telluric absorption. The resulting spectrum is plotted in Figure~\ref{fig:specnir}.

\section{Analysis} 
\label{sec:analysis}

\subsection{Photometric Evolution and Light Curve Modeling}
\label{subsec:photevo}

\begin{figure}
\center\includegraphics[width=0.8\linewidth]{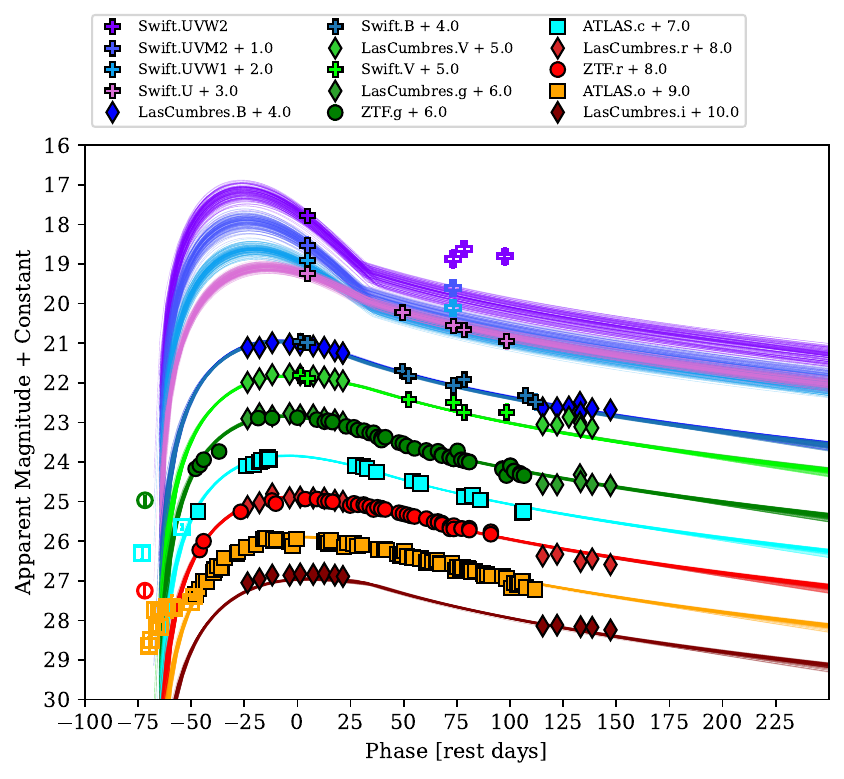}
\caption{Multi-band \sw{MOSFiT} model light curves for SN\,2024ahr. The model provides an excellent fit to the data over a span of about 200 days. The data points shown with open markers are not used in the fit (early bump, and UV points contaminated by host galaxy emission).}
\label{fig:mosfitlc}
\end{figure}

The UV and optical light curves of SN\,2024ahr are plotted in Figure~\ref{fig:lc}. The light curve of SN\,2024ahr is characterized by a gradual rise from discovery to a peak of $m_g \approx 17.1$ mag ($M_g \approx -21$) on 2024 April 3 (MJD = 60403.425, which we define as phase = 0 d), over a span of 78 days. After the peak, the light curve shows a shallow decline, reaching half brightness in g-band over 80 days. During the rising phase, the $g-r$ color gradually reddens from $\approx -0.3$ at discovery to $\approx 0.0$ at peak and then further reddens to $\approx 0.3$ at a phase of $+100$ days in the observed frame. The light curve evolution of SN\,2024ahr is typical of the SLSNe-I population \citep{2024MNRAS.535..471G}. Assuming a blackbody SED, we estimate a temperature of $\approx 10,500$ K and a photospheric radius of $\approx 4\times 10^{15}$ cm at the peak of the light curve.

\begin{table}[t!]
\caption{Parameter Posterior Values of the \sw{slsnni} Model from \sw{MOSFiT}.}
\centering
\begin{tabular}{lcc}
\toprule
Parameter & Prior & Posterior \\
\midrule
$P_{\mathrm{spin}}$ (ms) & Uniform &$3.32^{+0.30}_{-0.29}$ \\
$B$ ($\times 10^{14}$ G) & log  & $0.56^{+0.43}_{-0.11}$ \\
$M_{\mathrm{ej}}$ (M$_\odot)$ &  Uniform &$9.51^{+3.24}_{-2.55}$ \\
$v_{\mathrm{ej}}\,\,(\times 10^3\, {\rm km\,s}^{-1}$) & log &  $5.0^{+5.0}_{-2.5}$ \\
$t_{\rm exp}\,{\rm (days)}$ & Uniform &  $-19.5^{+0.9}_{-0.8}$ \\
$\log\, f_{\rm Ni}$ &  log & $-2.0^{+0.6}_{-0.7}$ \\
$\log\, n_{\rm H,host}$ & log & $20.8^{+0.1}_{-0.2}$ \\
$\lambda_{\mathrm{cutoff}}$ (\AA) & Uniform &$3468^{+199}_{-240}$ \\
$\alpha$ & Uniform & $1.76^{+0.44}_{-0.39}$ \\
$T_{\min}\,{\rm (K)}$ & Uniform & $9510^{+328}_{-449}$ \\
$M_{\mathrm{NS}}$ (M$_\odot)$ &  Uniform &$1.71^{+0.17}_{-0.16}$ \\
$\theta_{\rm PB}$ & Uniform & $0.93^{+0.49}_{-0.47}$ \\
$\log\, \sigma$ & log & $-1.13^{+0.02}_{-0.02}$\\
\bottomrule
\end{tabular}
\label{table:parameters}
\end{table} 

In the initial 20 days after discovery, SN\,2024ahr exhibits a clear ``bump'', about 2 mag fainter the mean peak in the high-cadence ATLAS $o$-band data. Such bumps have been observed in some previous SLSNe-I \citep{2016MNRAS.457L..79N, 2019MNRAS.487.2215A}, and the relative brightness of the SN\,2024ahr bump to its peak brightness is comparable to previous events. Such bumps have been speculated to arise from the shock breakout emission driven by the central engine in pre-expanded ejecta or due to interaction with a dense circumstellar medium confined in proximity to the progenitor. Given the paucity of data at other bands during this bump, we cannot determine its origin in detail.

We model the light curves of SN\,2024ahr using the Modular Open-Source Fitter for Transients \sw{MOSFiT} \citep{2017ascl.soft10006G} \sw{slsnni} model, following the approach and priors used in the large population study of \citet{2017ApJ...850...55N} and \citet{2024MNRAS.535..471G}.  We use a Gaussian prior on the ejecta velocity of $\sim 10^4$ km s$^{-1}$ based on the spectra. We used 200 MCMC walkers with 20,000 steps to ensure a thorough exploration of the parameter space and confirmed convergence using the Gelman-Rubin statistic, ensuring that all chains had a potential scale reduction factor below 1.2. We fit the multi-band light curves from phase $-55$ d onwards (i.e., neglecting the early-time bump, which the model cannot accommodate).  We also remove the late-time UV photometry (UVW1, UVW2, UVM2) at phase $\geq 75$ d) due to significant host contamination. The best-fit model light curves are shown in Figure~\ref{fig:mosfitlc}, providing a good match to the observed optical/UV light curves over a span of about 200 days. The posteriors of the model parameters are listed in Table~\ref{table:parameters}, and are typical of the SLSN population: $P_{\mathrm{spin}}\approx 3.3$ ms, $B\approx 5\times 10^{13}$ G, and $M_{\mathrm{ej}}\approx 10$ M$_{\odot}$.

\subsection{Optical Spectral Evolution and Local Metallicity} 
\label{sec:specevo}

We present the optical spectral evolution of SN\,2024ahr in Figure~\ref{fig:speceval}. The spectra before the peak are dominated by a blue continuum, with broad, weak absorption features of \ion{O}{2} with a blueshift velocity of $\approx$ 9746 kms$^{-1}$ (measured at phase = -26 d). In the near-peak phase,\ion{Fe}{2} + \ion{Mg}{2} features starts to emerge at $\approx 4400$\AA. In addition, the \ion{O}{1}$\lambda7774$ absorption feature starts to appear at the peak phase. We measure a velocity of $\approx 7700$ kms$^{-1}$ in near-peak (phase = +6 d). In the post-peak phase, \ion{O}{2} and \ion{Fe}{2}/ \ion{Mg}{2} absorption features start to weaken yet, remain present, and the \ion{O}{1}$\lambda7774$ feature gets stronger and becomes evident with a velocity of 6812  kms$^{-1}$ (at phase = +35d). Furthermore, \ion{Si}{2} $\lambda 6355$, \ion{C}{2} $\lambda 6580$ and $\lambda 7234$ and \ion{O}{1} $\lambda$7774 features emerge and become stronger over time. In the near-nebular phase (+118 d), we see the potential emergence of [\ion{O}{1}] $\lambda\lambda$6300, 6364 features, [\ion{Ca}{2}] $\lambda\lambda$7291, 7323 emission lines and likely emissions from \ion{Ca}{2} NIR triplet around 8500\AA. The spectral features seen in SN\,2024ahr spectrum are typical of SLSNe-I.

We detect a few host galaxy lines in the SN\,2024ahr spectra. The H$\alpha$, H$\beta$, H$\gamma$, \ion{Na}{1} D and [\ion{O}{3}] lines from the host are clearly visible in $\lambda$4500-$\lambda$7000 region of the spectra. We also detect \ion{Ca}{2} H, K lines at the rest wavelengths representing the Ca lines from the host. We use the host galaxy emission lines from the $+35$ d spectrum ($H\alpha$, $H\beta$ $H\gamma$, [\ion{O}{2}] and [\ion{O}{3}]) to determine the metalicity at the location of SN\,2024ahr, using the R23 diagnostic~\cite{1979MNRAS.189...95P, 1999ApJ...514..544K, 2006A&A...459...85N}.  We used 12 + log(O/H) -- [\ion{N}{2}]$\lambda 6584$/H$\alpha$ and 12 + log(O/H) -- [\ion{O}{3}]$\lambda 5007$/[\ion{N}{2}]$\lambda 6584$ metalicity correlations to estimate 12 + log(O/H) =$ 8.55 \pm 0.05$ and $8.64 \pm 0.09$ respectively. Both correlations rule out the lower R23 metalicity branch \citep{2006A&A...459...85N}. The value of $12 + {\rm log(O/H)}\approx 8.6$ is consistent with the range for SLSNe-I \citep{2014ApJ...787..138L, 2023MNRAS.524.3559C}.

\subsection{NIR Spectroscopic Analysis}

The NIR spectrum of SN\,2024ahr shown in Figure~\ref{fig:specnir} exhibits two broad features at $\approx 1.5\mu$m and $\approx 2.1 \mu$m. We identify the 1.5 $\mu$m feature as \ion{Mg}{1}$\,\lambda 1.575$ $\mu$m, and the 2.1 $\mu$m feature as a blend of \ion{Mg}{2}$\,\lambda 2.136$ $\mu$m and \ion{Co}{2}$\,\lambda 2.126$ $\mu$m; both sets of features are seen in stripped-envelope SNe, Type Ib and Ic \citep{2022ApJ...925..175S}.  We do not find clear evidence for the \ion{He}{1}$\, \lambda 2.058$ $\mu$m feature. We compare our spectrum to that of SN\,2019hge, which is the only SLSN-I to date that shows potential signs of helium in its NIR spectrum \citep{2020ApJ...902L...8Y}. The spectrum of SN\,2019hge shows a more convincing feature of \ion{He}{1}. with a blueshift of aout 6000 km s$^{-1}$), as well as a potential signature of \ion{Mg}{1}$\,\lambda 1.575$ $\mu$m, but it does not clearly exhibit the \ion{Mg}{2}/\ion{Co}{2} blend.

\begin{figure}[t!]
\center\includegraphics[width=0.8\linewidth]{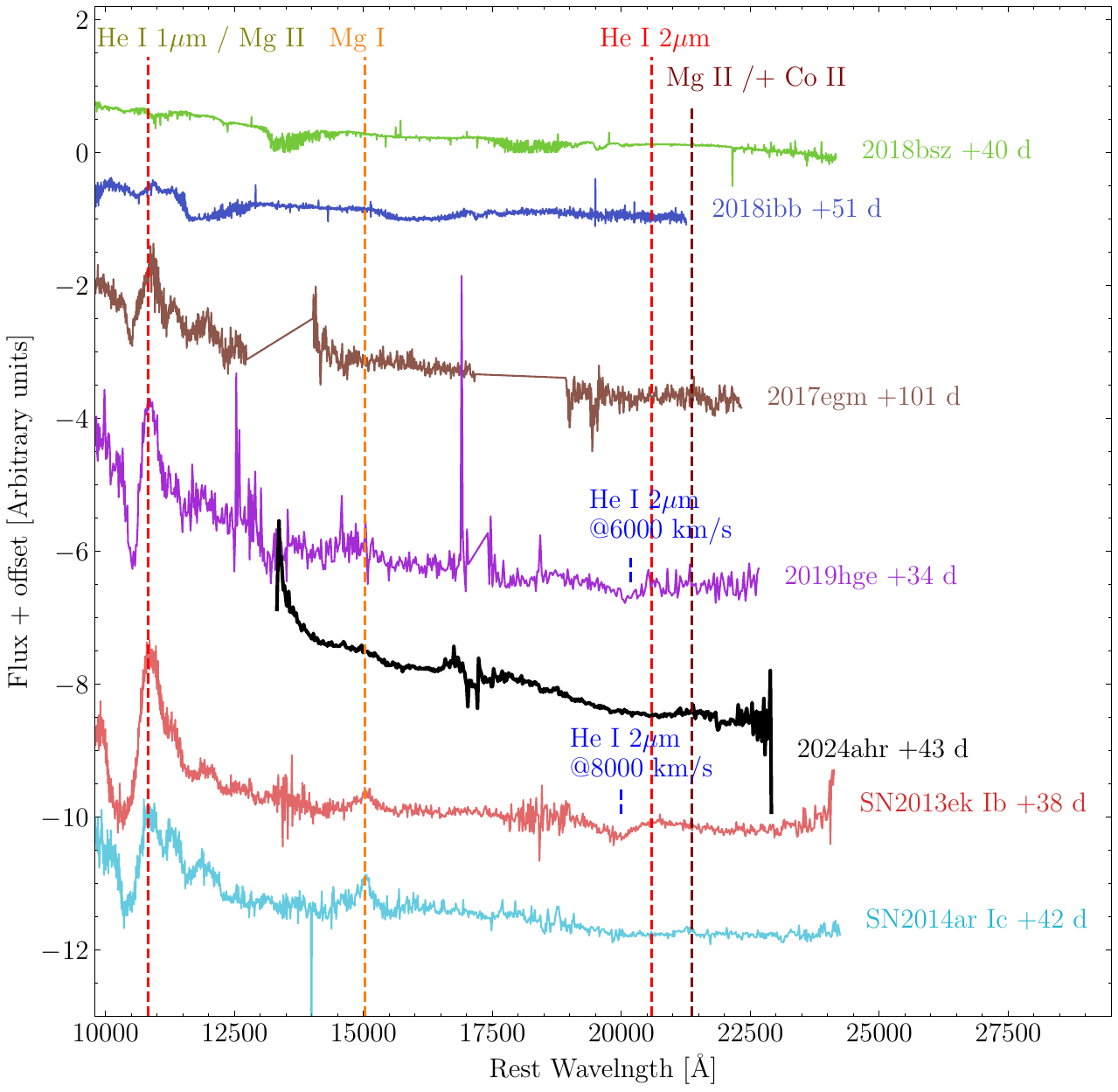}
\caption{NIR spectral comparison of SN\,2024ahr with all publicly available NIR spectra of SLSNe-I at a similar phase. We identify detections of \ion{Mg}{1} at $\approx 1.5 \mu m$ and \ion{Mg}{2}, \ion{Co}{2} feature at $\approx 2.1 \mu m$. All comparison SLSNe-I, except for SN2019hge, do not show any detection of the \ion{He}{1} 2$\mu$m feature in their NIR spectra. Data obtained from~\cite{2020ApJ...902L...8Y, 2022A&A...666A..30P, 2022ApJ...925..175S, 2023ApJ...949...23Z, 2024A&A...683A.223S}.}
\label{fig:speccompnir}
\end{figure}

A comparison of our SN\,2024ahr NIR spectrum with all publicly available NIR spectra of SLSNe-I at the nearest available rest-frame phase is shown in Figure~\ref{fig:speccompnir}. With the exception of SN\,2019hge, none of the other SLSNe-I show the presence of \ion{He}{1}$\,\lambda 2.058$ $\mu$m, or evidence for \ion{Mg}{1} or the \ion{Mg}{2}/\ion{Co}{2} blend.  We note that the previous events are mostly atypical of the general SLSN-I population: SN\,2019hge and SN\,2018bsz are at the low luminosity end of the population ($M_{r}\approx -19.6$ mag; \citealt{2020ApJ...902L...8Y}); SN\,2018ibb is a claimed pair-instability SN candidate \citep{2024A&A...683A.223S}; and SN\,2017egm has a claimed helium detection from optical spectra and the $1$ $\mu$m line \citep{2023ApJ...949...23Z}, but exhibits no clear detection of the 2$\mu$m line, which suggests that the previous claim is due to blending with other lines.

Also shown in Figure~\ref{fig:speccompnir} is a comparison with the NIR spectra of typical SNe Ib and Ic at a similar phase.  The comparison shows that SN\,2024ahr is more analogous to a SN Ic, exhibiting 
\ion{Mg}{1} and \ion{Mg}{2}/\ion{Co}{2}, but with higher velocities, while being distinguished from a SN Ib by lacking the clear \ion{He}{1}$\,\lambda 2.058$ $\mu$m line.

To assess a limit on the mass of helium in the outer layers of the ejecta from the lack of \ion{He}{1}$\,\lambda 2.058$ $\mu$m detection, we model the spectrum of SN\,2024ahr using the \sw{TARDIS} code \citep{2014MNRAS.440..387K}, which provides rapid spectral modeling using parameters such as composition, density profile, velocity, luminosity, and temperature as input. We use a model similar to that presented in \citet{2021ApJ...908..150W} with the luminosity, ejecta mass, velocity, temperature, density, and abundances adjusted to match the photometric and spectroscopic properties of SN\,2024ahr. We then add varying amounts of helium (0, 0.025, and 0.05 M$_\odot$) to the outer layers of the ejecta to estimate the amount of helium that can be present in SN\,2024ahr at the time of our NIR spectrum. 

\begin{figure}[t!]
\center\includegraphics[width=0.9\linewidth]{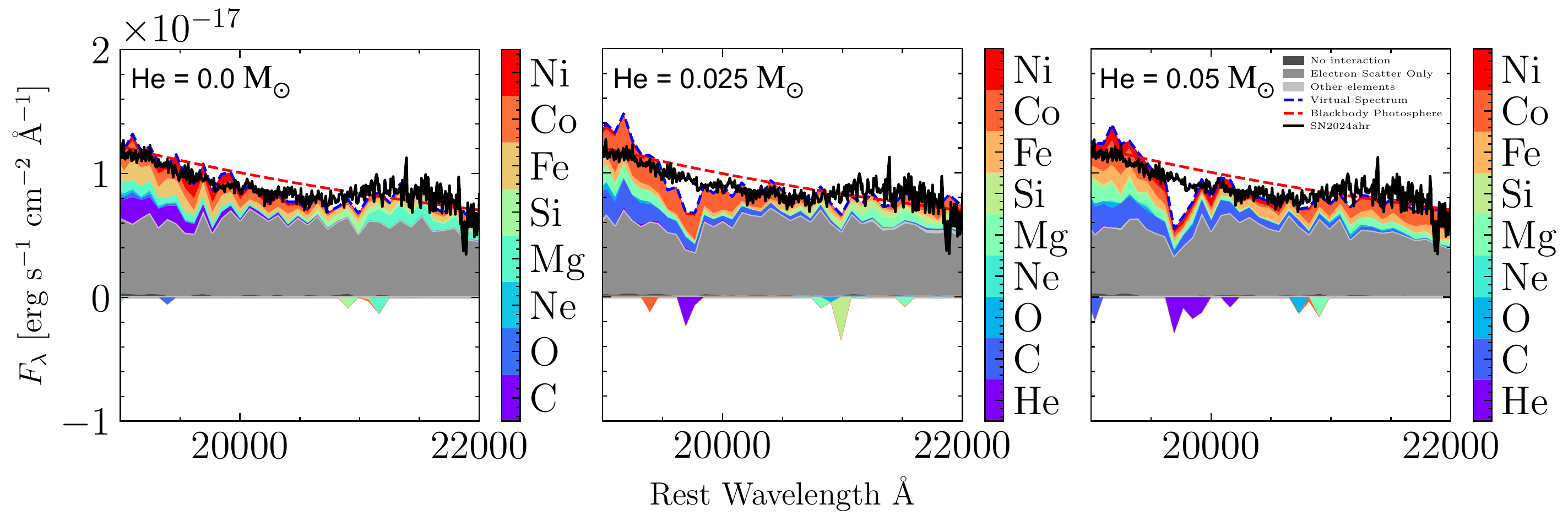}
\caption{\sw{TARDIS} model spectra with helium masses of 0, 0.025, and 0.05 M$_\odot$, compared to our spectrum of SN\,2024ahr near the \ion{He}{1}$\,\lambda 2.058$ $\mu$m line. The contribution of various element is represented by different colors labeled in the color bar.  We find that a helium mass of 0.05 M$_\odot$ would have produced a notable absorption feature in the spectrum, and use this value as a rough upper limit on the helium mass in the outer layers of the ejecta.}
\label{fig:spectardis}
\end{figure}

A comparison of the our SN\,2024ahr spectrum to the  \sw{TARDIS} synthetic spectra around the location of the \ion{He}{1}$\,\lambda 2.058$ $\mu$m line is shown in Figure~\ref{fig:spectardis}. The model provides an overall good fit to the observed spectrum, including the observed feature due to \ion{Mg}{2}/\ion{Co}{2}.  We find that a helium mass of $\sim 0.05$ M$_\odot$ would have produced a notable absorption feature that is not present in the data, and we therefore use this value as a conservative upper limit on the helium mass in the outer ejecta layers of SN\,2024ahr.

\section{Conclusions}
\label{sec:conclusion}

We presented detailed photometric and spectroscopic observations of the SLSN-I SN\,2024ahr, and determined its precise redshift to $z=0.0861$, making it one of the nearest SLSN-I to date.  The optical/UV light curves, optical spectra, and inferred model parameters of SN\,2024ahr are all typical of the SLSN-I population, and specifically with the median peak absolute magnitude of $\sim -21$.

Taking advantage of the low redshift, we obtained a high signal-to-noise ratio NIR spectrum to robustly explore for presence of helium via the \ion{He}{1}$\,\lambda 2.058$ $\mu$m line.  The spectrum reveals broad features of \ion{Mg}{1}, \ion{Mg}{2} + \ion{Co}{2} which are typical of Type Ic SNe, but with hogher expansion velocities. We do not detect absorption at the location of the \ion{He}{1}$\lambda 2.058$ $\mu$m line, and using spectral modeling, we place a rough upper limit of $\sim 0.05$ M$_\odot$ on the mass of helium present in the outer ejecta.  This suggests that the progenitor of SN\,2024ahr was stripped of both hydrogen and helium (i.e., Ic-like).  Detailed NIR ground-based spectroscopy in progress will determine whether any other SLSNe-I at $z\lesssim 0.1$ exhibit robust evidence for helium, and future observations with JWST will provide such insight over a broader redshift range and to fainter magnitudes.

\begin{acknowledgments}

We are grateful to Kali Salmas, Alejandra Milone and  Benjamin Weiner for scheduling the MMT Binospec observations and Yuri Beletsky for performing the Magellan LDSS-3 observations.

The Berger Time-Domain research group at Harvard is supported by the NSF and NASA grants. The LCO supernova group is supported by NSF grants AST-1911151 and AST-1911225.

Observations reported here were obtained at the MMT Observatory, a joint facility of the Smithsonian Institution and the University of Arizona. This paper uses data products produced by the OIR Telescope Data Center, supported by the Smithsonian Astrophysical Observatory.
This work makes use of observations from the Las Cumbres Observatory global telescope network. The authors wish to recognize and acknowledge the very significant cultural role and reverence that the summit of Haleakalā has always had within the indigenous Hawaiian community. We are most fortunate to have the opportunity to conduct observations from the mountain.

We thank the support of the staff at the Neil Gehrels Swift Observatory.

This research made use of \sw{PypeIt},\footnote{\url{https://pypeit.readthedocs.io/en/latest/}}
a Python package for semi-automated reduction of astronomical slit-based spectroscopy
\citep{pypeit:joss_pub, pypeit:zenodo}.

This work has made use of data from the Zwicky Transient Facility (ZTF). ZTF is supported by NSF grant No. AST- 1440341 and a collaboration including Caltech, IPAC, the Weizmann Institute for Science, the Oskar Klein Center at Stockholm University, the University of Maryland, the University of Washington, Deutsches Elektronen-Synchrotron and Humboldt University, Los Alamos National Laboratories, the TANGO Consortium of Taiwan, the University of Wisconsin–Milwaukee, and Lawrence Berkeley National Laboratories. Operations are conducted by COO, IPAC, and UW. The ZTF forced-photometry service was funded under the Heising-Simons Foundation grant No. 12540303 (PI: Graham).

This work has made use of data from the Asteroid Terrestrial-impact Last Alert System (ATLAS) project. The Asteroid Terrestrial-impact Last Alert System (ATLAS) project is primarily funded to search for near-earth asteroids through NASA grants NN12AR55G, 80NSSC18K0284, and 80NSSC18K1575; byproducts of the NEO search include images and catalogs from the survey area. This work was partially funded by Kepler/K2 grant J1944/80NSSC19K0112 and HST GO-15889, and STFC grants ST/T000198/1 and ST/S006109/1. The ATLAS science products have been made possible through the contributions of the University of Hawaii Institute for Astronomy, the Queen’s University Belfast, the Space Telescope Science Institute, the South African Astronomical Observatory, and The Millennium Institute of Astrophysics (MAS), Chile.
This research made use of \textsc{tardis}, a community-developed software package for spectral synthesis in supernovae \citep{2014MNRAS.440..387K, kerzendorf_2024_13929578}. The development of \textsc{tardis} received support from GitHub, the Google Summer of Code initiative, and from ESA's Summer of Code in Space program. \textsc{tardis} is a fiscally sponsored project of NumFOCUS. \textsc{tardis} makes extensive use of Astropy and Pyne.

This research has made use of the NASA Astrophysics Data System (ADS), the NASA/IPAC Extragalactic Database (NED), and NASA/IPAC Infrared Science Archive (IRSA, which is funded by NASA and operated by the California Institute of Technology) and IRAF (which is distributed by the National Optical Astronomy Observatory, NOAO, operated by the Association of Universities for Research in Astronomy, AURA, Inc., under cooperative agreement with the NSF).

TNS is supported by funding from the Weizmann Institute of Science, as well as grants from the Israeli Institute for Advanced Studies and the European Union via ERC grant No. 725161.

\end{acknowledgments}

\vspace{5mm}
\facilities{ATLAS, LCO, Swift(XRT and UVOT) and ZTF, MMT}

\software{astropy~\citep{2013A&A...558A..33A,2018AJ....156..123A, 2022ApJ...935..167A}, \sw{SExtractor}~\citep{1996A&AS..117..393B} \sw{NumPy}~\citep{oliphant2015guide}, \sw{photutils}~\citep{2022zndo...6825092B}, \sw{PyRAF}~\citep{2012ascl.soft07011S}, \sw{SciPy}~\citep{2020SciPy-NMeth} and \sw{MOSFiT}~\citep{2017ascl.soft10006G}}
\newpage

\appendix
\section{Photometry}

\begin{longtable}{|c|c|c|c|c|} 
\hline
\hline
MJD &  Filter & Mag System & Magnitude  $\pm$  e\_magnitude (AB) & Telescope \\
\hline
\hline
60324.08414 & c & AB & $ 19.31 \pm 0.17 $ & ATLAS \\ 
60324.63717 & o & AB & $ >19.71 \pm 0.28 $ & ATLAS \\ 
60325.48944 & r & AB & $ 19.25 \pm 0.17 $ & ZTF \\ 
60325.52751 & g & AB & $ 18.97 \pm 0.10 $ & ZTF \\ 
60327.64583 & o & AB & $ 19.64 \pm 0.2 $ & ATLAS \\ 
60328.65230 & o & AB & $ 19.5 \pm 0.12 $ & ATLAS \\ 
60330.61142 & o & AB & $ 18.73 \pm 0.13 $ & ATLAS \\ 
60331.35503 & o & AB & $ 19.18 \pm 0.12 $ & ATLAS \\ 
60332.58525 & o & AB & $ >18.8 \pm 0.23 $ & ATLAS \\ 
60333.09649 & o & AB & $ 19.17 \pm 0.13 $ & ATLAS \\ 
60336.57452 & o & AB & $ 18.67 \pm 0.17 $ & ATLAS \\ 
60339.30233 & o & AB & $ >18.65 \pm 0.28 $ & ATLAS \\ 
60340.63145 & o & AB & $ 18.7 \pm 0.19 $ & ATLAS \\ 
60344.58362 & o & AB & $ 18.64 \pm 0.07 $ & ATLAS \\ 
60348.61786 & c & AB & $ 18.6 \pm 0.06 $ & ATLAS \\ 
60349.11079 & o & AB & $ 18.54 \pm 0.06 $ & ATLAS \\ 
60350.59033 & o & AB & $ 18.4 \pm 0.05 $ & ATLAS \\ 
60351.30575 & o & AB & $ 18.35 \pm 0.05 $ & ATLAS \\ 
60351.40104 & g & AB & $ 18.17 \pm 0.08 $ & ZTF \\ 
60352.59256 & c & AB & $ 18.26 \pm 0.04 $ & ATLAS \\ 
60353.07085 & o & AB & $ 18.21 \pm 0.04 $ & ATLAS \\ 
60353.46939 & r & AB & $ 18.23 \pm 0.080 $ & ZTF \\ 
60353.50177 & g & AB & $ 18.08 \pm 0.080 $ & ZTF \\ 
60355.31720 & o & AB & $ 18.02 \pm 0.04 $ & ATLAS \\ 
60355.47949 & r & AB & $ 18.0 \pm 0.070 $ & ZTF \\ 
60355.52082 & g & AB & $ 17.94 \pm 0.05 $ & ZTF \\ 
60357.06662 & o & AB & $ 18.02 \pm 0.03 $ & ATLAS \\ 
60360.57107 & o & AB & $ 17.79 \pm 0.040 $ & ATLAS \\ 
60361.05397 & o & AB & $ 17.72 \pm 0.03 $ & ATLAS \\ 
60362.31309 & o & AB & $ 17.62 \pm 0.05 $ & ATLAS \\ 
60363.40083 & g & AB & $ 17.74 \pm 0.15 $ & ZTF \\ 
60364.47799 & o & AB & $ 17.66 \pm 0.12 $ & ATLAS \\ 
60366.28186 & o & AB & $ 17.42 \pm 0.05 $ & ATLAS \\ 
60372.54168 & o & AB & $ 17.31 \pm 0.03 $ & ATLAS \\ 
60373.08527 & o & AB & $ 17.28 \pm 0.09 $ & ATLAS \\ 
60374.45652 & r & AB & $ 17.26 \pm 0.05 $ & ZTF \\ 
60377.11317 & c & AB & $ 17.1 \pm 0.02 $ & ATLAS \\ 
60377.28392 & o & AB & $ 17.16 \pm 0.02 $ & ATLAS \\ 
60377.92168 & B &  Vega & $ 17.11 \pm 0.01 $ & LasCumbres \\ 
60377.92451 & V &  Vega & $ 16.99 \pm 0.02 $ & LasCumbres \\ 
60377.92642 & g & AB & $ 16.91 \pm 0.02 $ & LasCumbres \\ 
60377.92924 & r & AB & $ 17.12 \pm 0.02 $ & LasCumbres \\ 
60377.93117 & i & AB & $ 17.05 \pm 0.03 $ & LasCumbres \\ 
60380.96319 & c & AB & $ 17.06 \pm 0.02 $ & ATLAS \\ 
60382.26828 & o & AB & $ 17.09 \pm 0.02 $ & ATLAS \\ 
60383.39547 & g & AB & $ 16.89 \pm 0.06 $ & ZTF \\ 
60384.01817 & B &  Vega & $ 17.11 \pm 0.02 $ & LasCumbres \\ 
60384.02100 & V &  Vega & $ 16.9 \pm 0.02 $ & LasCumbres \\ 
60384.02122 & c & AB & $ 17.0 \pm 0.020 $ & ATLAS \\ 
60384.02290 & g & AB & $ 16.78 \pm 0.02 $ & LasCumbres \\ 
60384.02576 & r & AB & $ 17.03 \pm 0.02 $ & LasCumbres \\ 
60384.02767 & i & AB & $ 16.95 \pm 0.02 $ & LasCumbres \\ 
60385.08850 & c & AB & $ 16.97 \pm 0.01 $ & ATLAS \\ 
60386.25692 & o & AB & $ 16.95 \pm 0.02 $ & ATLAS \\ 
60387.24077 & o & AB & $ 16.93 \pm 0.02 $ & ATLAS \\ 
60387.99893 & c & AB & $ 16.88 \pm 0.03 $ & ATLAS \\ 
60388.01445 & c & AB & $ 16.91 \pm 0.02 $ & ATLAS \\ 
60389.01396 & c & AB & $ 16.93 \pm 0.02 $ & ATLAS \\ 
60390.39617 & r & AB & $ 16.97 \pm 0.040 $ & ZTF \\ 
60390.46168 & g & AB & $ 16.89 \pm 0.040 $ & ZTF \\ 
60390.55718 & B &  Vega & $ 16.99 \pm 0.02 $ & LasCumbres \\ 
60390.56002 & V &  Vega & $ 16.79 \pm 0.02 $ & LasCumbres \\ 
60390.56196 & g & AB & $ 16.84 \pm 0.01 $ & LasCumbres \\ 
60390.56481 & r & AB & $ 16.81 \pm 0.02 $ & LasCumbres \\ 
60390.56674 & i & AB & $ 16.86 \pm 0.02 $ & LasCumbres \\ 
60390.99709 & o & AB & $ 16.98 \pm 0.02 $ & ATLAS \\ 
60391.17561 & o & AB & $ 16.96 \pm 0.02 $ & ATLAS \\ 
60392.36505 & r & AB & $ 17.05 \pm 0.120 $ & ZTF \\ 
60392.43633 & o & AB & $ 16.97 \pm 0.04 $ & ATLAS \\ 
60399.35003 & o & AB & $ 17.01 \pm 0.02 $ & ATLAS \\ 
60399.42088 & B &  Vega & $ 17.02 \pm 0.02 $ & LasCumbres \\ 
60399.42371 & V &  Vega & $ 16.78 \pm 0.03 $ & LasCumbres \\ 
60399.42561 & g & AB & $ 16.78 \pm 0.02 $ & LasCumbres \\ 
60399.42842 & r & AB & $ 16.9 \pm 0.03 $ & LasCumbres \\ 
60399.43035 & i & AB & $ 16.86 \pm 0.04 $ & LasCumbres \\ 
60400.44570 & o & AB & $ 16.97 \pm 0.020 $ & ATLAS \\ 
60401.00885 & o & AB & $ 17.12 \pm 0.08 $ & ATLAS \\ 
60402.95293 & o & AB & $ 16.95 \pm 0.01 $ & ATLAS \\ 
60403.42513 & g & AB & $ 16.88 \pm 0.040 $ & ZTF \\ 
60404.83115 & B &  Vega & $ 17.04 \pm 0.02 $ & LasCumbres \\ 
60404.83401 & V &  Vega & $ 16.8 \pm 0.02 $ & LasCumbres \\ 
60404.83596 & g & AB & $ 16.82 \pm 0.02 $ & LasCumbres \\ 
60404.83882 & r & AB & $ 16.89 \pm 0.02 $ & LasCumbres \\ 
60404.84073 & i & AB & $ 16.83 \pm 0.02 $ & LasCumbres \\ 
60404.87142 & B &  Vega & $ 16.97 \pm 0.070 $ & Swift \\ 
60404.87947 & V &  Vega & $ 16.81 \pm 0.09 $ & Swift \\ 
60407.38877 & r & AB & $ 16.94 \pm 0.050 $ & ZTF \\ 
60408.49503 & UVW1 &  Vega & $ 16.9 \pm 0.12 $ & Swift \\ 
60408.49590 & U &  Vega & $ 16.25 \pm 0.09 $ & Swift \\ 
60408.49677 & B &  Vega & $ 16.99 \pm 0.09 $ & Swift \\ 
60408.50012 & UVW2 &  Vega & $ 17.79 \pm 0.15 $ & Swift \\ 
60408.50100 & V &  Vega & $ 16.89 \pm 0.13 $ & Swift \\ 
60408.50279 & UVM2 &  Vega & $ 17.54 \pm 0.21 $ & Swift \\ 
60411.49318 & B &  Vega & $ 17.07 \pm 0.01 $ & LasCumbres \\ 
60411.49605 & V &  Vega & $ 16.83 \pm 0.02 $ & LasCumbres \\ 
60411.49802 & g & AB & $ 16.81 \pm 0.01 $ & LasCumbres \\ 
60411.50088 & r & AB & $ 16.91 \pm 0.02 $ & LasCumbres \\ 
60411.50284 & i & AB & $ 16.84 \pm 0.02 $ & LasCumbres \\ 
60413.33685 & g & AB & $ 16.93 \pm 0.040 $ & ZTF \\ 
60413.37721 & r & AB & $ 16.94 \pm 0.030 $ & ZTF \\ 
60417.02692 & o & AB & $ 17.01 \pm 0.02 $ & ATLAS \\ 
60417.11246 & B &  Vega & $ 17.1 \pm 0.02 $ & LasCumbres \\ 
60417.11529 & V &  Vega & $ 16.85 \pm 0.02 $ & LasCumbres \\ 
60417.11721 & g & AB & $ 16.89 \pm 0.02 $ & LasCumbres \\ 
60417.12007 & r & AB & $ 16.91 \pm 0.02 $ & LasCumbres \\ 
60417.12196 & i & AB & $ 16.82 \pm 0.03 $ & LasCumbres \\ 
60417.35166 & g & AB & $ 16.98 \pm 0.040 $ & ZTF \\ 
60417.35888 & r & AB & $ 16.99 \pm 0.040 $ & ZTF \\ 
60418.25190 & o & AB & $ 16.98 \pm 0.02 $ & ATLAS \\ 
60419.01736 & o & AB & $ 17.05 \pm 0.03 $ & ATLAS \\ 
60419.26964 & r & AB & $ 16.99 \pm 0.040 $ & ZTF \\ 
60419.31528 & g & AB & $ 16.96 \pm 0.03 $ & ZTF \\ 
60420.60103 & o & AB & $ 16.98 \pm 0.02 $ & ATLAS \\ 
60421.32247 & g & AB & $ 16.99 \pm 0.040 $ & ZTF \\ 
60421.38245 & r & AB & $ 17.01 \pm 0.05 $ & ZTF \\ 
60422.55031 & B &  Vega & $ 17.2 \pm 0.02 $ & LasCumbres \\ 
60422.55317 & V &  Vega & $ 16.91 \pm 0.02 $ & LasCumbres \\ 
60422.55512 & g & AB & $ 16.94 \pm 0.02 $ & LasCumbres \\ 
60422.55800 & r & AB & $ 16.97 \pm 0.02 $ & LasCumbres \\ 
60422.55997 & i & AB & $ 16.86 \pm 0.02 $ & LasCumbres \\ 
60426.60778 & B &  Vega & $ 17.26 \pm 0.03 $ & LasCumbres \\ 
60426.66139 & B &  Vega & $ 17.25 \pm 0.02 $ & LasCumbres \\ 
60426.66561 & V &  Vega & $ 16.95 \pm 0.02 $ & LasCumbres \\ 
60426.66754 & g & AB & $ 16.99 \pm 0.02 $ & LasCumbres \\ 
60426.67038 & r & AB & $ 16.99 \pm 0.02 $ & LasCumbres \\ 
60426.67228 & i & AB & $ 16.89 \pm 0.03 $ & LasCumbres \\ 
60427.21691 & o & AB & $ 17.15 \pm 0.07 $ & ATLAS \\ 
60428.38228 & g & AB & $ 17.1 \pm 0.050 $ & ZTF \\ 
60429.03034 & o & AB & $ 17.05 \pm 0.02 $ & ATLAS \\ 
60430.38148 & r & AB & $ 17.1 \pm 0.05 $ & ZTF \\ 
60431.99050 & o & AB & $ 17.05 \pm 0.02 $ & ATLAS \\ 
60432.00598 & o & AB & $ 17.06 \pm 0.05 $ & ATLAS \\ 
60432.31221 & g & AB & $ 17.13 \pm 0.040 $ & ZTF \\ 
60432.33711 & r & AB & $ 17.04 \pm 0.06 $ & ZTF \\ 
60432.96333 & c & AB & $ 17.09 \pm 0.02 $ & ATLAS \\ 
60434.21322 & o & AB & $ 17.12 \pm 0.02 $ & ATLAS \\ 
60434.33249 & g & AB & $ 17.19 \pm 0.03 $ & ZTF \\ 
60434.33771 & r & AB & $ 17.08 \pm 0.040 $ & ZTF \\ 
60436.44816 & o & AB & $ 17.1 \pm 0.02 $ & ATLAS \\ 
60436.96603 & c & AB & $ 17.12 \pm 0.02 $ & ATLAS \\ 
60437.25648 & r & AB & $ 17.07 \pm 0.04 $ & ZTF \\ 
60437.44613 & g & AB & $ 17.21 \pm 0.060 $ & ZTF \\ 
60438.26602 & r & AB & $ 17.11 \pm 0.050 $ & ZTF \\ 
60438.92011 & c & AB & $ 17.17 \pm 0.02 $ & ATLAS \\ 
60440.31531 & g & AB & $ 17.27 \pm 0.040 $ & ZTF \\ 
60440.36323 & r & AB & $ 17.13 \pm 0.040 $ & ZTF \\ 
60442.27101 & r & AB & $ 17.2 \pm 0.03 $ & ZTF \\ 
60442.31830 & g & AB & $ 17.26 \pm 0.030 $ & ZTF \\ 
60443.92262 & c & AB & $ 17.26 \pm 0.02 $ & ATLAS \\ 
60444.27338 & r & AB & $ 17.14 \pm 0.050 $ & ZTF \\ 
60444.32509 & g & AB & $ 17.37 \pm 0.050 $ & ZTF \\ 
60444.76937 & o & AB & $ 17.22 \pm 0.04 $ & ATLAS \\ 
60446.30615 & r & AB & $ 17.17 \pm 0.05 $ & ZTF \\ 
60446.30615 & r & AB & $ 17.19 \pm 0.02 $ & ZTF \\ 
60446.35646 & g & AB & $ 17.39 \pm 0.050 $ & ZTF \\ 
60446.35646 & g & AB & $ 17.45 \pm 0.03 $ & ZTF \\ 
60447.91065 & o & AB & $ 17.21 \pm 0.03 $ & ATLAS \\ 
60448.27046 & r & AB & $ 17.2 \pm 0.050 $ & ZTF \\ 
60448.36916 & g & AB & $ 17.38 \pm 0.05 $ & ZTF \\ 
60448.82955 & o & AB & $ 17.22 \pm 0.030 $ & ATLAS \\ 
60454.41774 & o & AB & $ 17.33 \pm 0.05 $ & ATLAS \\ 
60455.23133 & g & AB & $ 17.51 \pm 0.040 $ & ZTF \\ 
60455.29560 & r & AB & $ 17.28 \pm 0.08 $ & ZTF \\ 
60455.98819 & o & AB & $ 17.36 \pm 0.04 $ & ATLAS \\ 
60456.91827 & o & AB & $ 17.33 \pm 0.020 $ & ATLAS \\ 
60457.25102 & g & AB & $ 17.53 \pm 0.040 $ & ZTF \\ 
60457.26601 & U &  Vega & $ 17.23 \pm 0.16 $ & Swift \\ 
60457.26655 & B &  Vega & $ 17.7 \pm 0.15 $ & Swift \\ 
60457.31492 & r & AB & $ 17.31 \pm 0.05 $ & ZTF \\ 
60458.22738 & o & AB & $ 17.25 \pm 0.03 $ & ATLAS \\ 
60459.23141 & g & AB & $ 17.6 \pm 0.05 $ & ZTF \\ 
60459.35644 & r & AB & $ 17.33 \pm 0.04 $ & ZTF \\ 
60459.47113 & o & AB & $ 17.38 \pm 0.05 $ & ATLAS \\ 
60460.36116 & B &  Vega & $ 17.83 \pm 0.14 $ & Swift \\ 
60460.36531 & V &  Vega & $ 17.43 \pm 0.19 $ & Swift \\ 
60461.29211 & r & AB & $ 17.36 \pm 0.040 $ & ZTF \\ 
60462.13953 & c & AB & $ 17.48 \pm 0.020 $ & ATLAS \\ 
60462.35592 & o & AB & $ 17.41 \pm 0.02 $ & ATLAS \\ 
60463.23333 & r & AB & $ 17.38 \pm 0.050 $ & ZTF \\ 
60463.27476 & g & AB & $ 17.66 \pm 0.05 $ & ZTF \\ 
60466.34191 & o & AB & $ 17.43 \pm 0.02 $ & ATLAS \\ 
60466.37715 & c & AB & $ 17.55 \pm 0.03 $ & ATLAS \\ 
60469.22962 & g & AB & $ 17.71 \pm 0.05 $ & ZTF \\ 
60469.27388 & r & AB & $ 17.43 \pm 0.040 $ & ZTF \\ 
60469.33921 & o & AB & $ 17.54 \pm 0.020 $ & ATLAS \\ 
60470.36083 & o & AB & $ 17.49 \pm 0.02 $ & ATLAS \\ 
60471.22924 & g & AB & $ 17.77 \pm 0.05 $ & ZTF \\ 
60472.40384 & o & AB & $ 17.52 \pm 0.03 $ & ATLAS \\ 
60473.24841 & r & AB & $ 17.52 \pm 0.050 $ & ZTF \\ 
60474.62768 & o & AB & $ 17.47 \pm 0.02 $ & ATLAS \\ 
60475.21093 & g & AB & $ 17.73 \pm 0.06 $ & ZTF \\ 
60475.23213 & r & AB & $ 17.52 \pm 0.050 $ & ZTF \\ 
60475.23213 & r & AB & $ 17.5 \pm 0.02 $ & ZTF \\ 
60476.57139 & o & AB & $ 17.57 \pm 0.03 $ & ATLAS \\ 
60477.20998 & g & AB & $ 17.82 \pm 0.07 $ & ZTF \\ 
60477.20998 & g & AB & $ 17.84 \pm 0.04 $ & ZTF \\ 
60477.32791 & r & AB & $ 17.54 \pm 0.070 $ & ZTF \\ 
60477.32791 & r & AB & $ 17.58 \pm 0.04 $ & ZTF \\ 
60481.23105 & g & AB & $ 17.87 \pm 0.08 $ & ZTF \\ 
60481.23105 & g & AB & $ 17.89 \pm 0.06 $ & ZTF \\ 
60481.30275 & r & AB & $ 17.64 \pm 0.07 $ & ZTF \\ 
60481.30275 & r & AB & $ 17.69 \pm 0.05 $ & ZTF \\ 
60482.34075 & o & AB & $ 17.57 \pm 0.050 $ & ATLAS \\ 
60483.03957 & UVW1 &  Vega & $ 18.12 \pm 0.17 $ & Swift \\ 
60483.04061 & U &  Vega & $ 17.56 \pm 0.14 $ & Swift \\ 
60483.04166 & B &  Vega & $ 18.07 \pm 0.13 $ & Swift \\ 
60483.04671 & UVW2 &  Vega & $ 18.87 \pm 0.22 $ & Swift \\ 
60483.04776 & V &  Vega & $ 17.5 \pm 0.16 $ & Swift \\ 
60483.05557 & UVM2 &  Vega & $ 18.6 \pm 0.18 $ & Swift \\ 
60483.24998 & g & AB & $ 17.92 \pm 0.08 $ & ZTF \\ 
60483.24998 & g & AB & $ 17.92 \pm 0.06 $ & ZTF \\ 
60483.27690 & r & AB & $ 17.69 \pm 0.06 $ & ZTF \\ 
60483.27690 & r & AB & $ 17.61 \pm 0.03 $ & ZTF \\ 
60483.90807 & o & AB & $ 17.74 \pm 0.07 $ & ATLAS \\ 
60484.88288 & o & AB & $ 17.66 \pm 0.07 $ & ATLAS \\ 
60485.23256 & g & AB & $ 17.72 \pm 0.050 $ & ZTF \\ 
60486.64395 & o & AB & $ 17.68 \pm 0.03 $ & ATLAS \\ 
60487.21367 & g & AB & $ 17.95 \pm 0.06 $ & ZTF \\ 
60487.21367 & g & AB & $ 17.95 \pm 0.02 $ & ZTF \\ 
60487.29531 & r & AB & $ 17.68 \pm 0.06 $ & ZTF \\ 
60487.29531 & r & AB & $ 17.68 \pm 0.030 $ & ZTF \\ 
60487.88439 & o & AB & $ 17.66 \pm 0.03 $ & ATLAS \\ 
60488.56850 & U &  Vega & $ 17.67 \pm 0.15 $ & Swift \\ 
60488.56947 & B &  Vega & $ 17.93 \pm 0.13 $ & Swift \\ 
60488.57408 & UVW2 &  Vega & $ 18.63 \pm 0.19 $ & Swift \\ 
60488.57505 & V &  Vega & $ 17.75 \pm 0.20 $ & Swift \\ 
60488.88209 & c & AB & $ 17.88 \pm 0.03 $ & ATLAS \\ 
60489.21492 & g & AB & $ 17.97 \pm 0.06 $ & ZTF \\ 
60489.21492 & g & AB & $ 17.98 \pm 0.030 $ & ZTF \\ 
60490.15739 & o & AB & $ 17.7 \pm 0.03 $ & ATLAS \\ 
60491.07838 & o & AB & $ 17.7 \pm 0.03 $ & ATLAS \\ 
60491.22653 & g & AB & $ 18.0 \pm 0.06 $ & ZTF \\ 
60491.22653 & g & AB & $ 18.0 \pm 0.03 $ & ZTF \\ 
60491.25539 & r & AB & $ 17.72 \pm 0.050 $ & ZTF \\ 
60491.25539 & r & AB & $ 17.67 \pm 0.02 $ & ZTF \\ 
60492.38837 & o & AB & $ 17.77 \pm 0.03 $ & ATLAS \\ 
60492.81733 & c & AB & $ 17.85 \pm 0.03 $ & ATLAS \\ 
60494.19315 & o & AB & $ 17.75 \pm 0.02 $ & ATLAS \\ 
60496.80960 & c & AB & $ 17.96 \pm 0.03 $ & ATLAS \\ 
60498.19322 & o & AB & $ 17.81 \pm 0.020 $ & ATLAS \\ 
60499.05855 & o & AB & $ 17.87 \pm 0.03 $ & ATLAS \\ 
60500.30509 & o & AB & $ 17.85 \pm 0.03 $ & ATLAS \\ 
60502.24168 & r & AB & $ 17.83 \pm 0.06 $ & ZTF \\ 
60502.24168 & r & AB & $ 17.76 \pm 0.03 $ & ZTF \\ 
60502.27800 & o & AB & $ 17.88 \pm 0.03 $ & ATLAS \\ 
60508.20521 & g & AB & $ 18.17 \pm 0.12 $ & ZTF \\ 
60508.20521 & g & AB & $ 18.19 \pm 0.09 $ & ZTF \\ 
60508.32324 & o & AB & $ 17.88 \pm 0.090 $ & ATLAS \\ 
60509.33890 & UVW2 &  Vega & $ 18.81 \pm 0.20 $ & Swift \\ 
60510.21934 & o & AB & $ 17.95 \pm 0.05 $ & ATLAS \\ 
60510.22771 & g & AB & $ 18.34 \pm 0.15 $ & ZTF \\ 
60510.38450 & U &  Vega & $ 17.95 \pm 0.21 $ & Swift \\ 
60510.39098 & V &  Vega & $ 17.75 \pm 0.21 $ & Swift \\ 
60512.22896 & g & AB & $ 18.1 \pm 0.14 $ & ZTF \\ 
60512.22896 & g & AB & $ 18.1 \pm 0.1 $ & ZTF \\ 
60512.85743 & o & AB & $ 18.18 \pm 0.09 $ & ATLAS \\ 
60514.19013 & g & AB & $ 18.23 \pm 0.10 $ & ZTF \\ 
60514.19013 & g & AB & $ 18.23 \pm 0.06 $ & ZTF \\ 
60514.22316 & o & AB & $ 18.06 \pm 0.08 $ & ATLAS \\ 
60515.85070 & o & AB & $ 18.11 \pm 0.08 $ & ATLAS \\ 
60516.36757 & V &  Vega & $ 17.42 \pm 0.21 $ & Swift \\ 
60516.79299 & o & AB & $ 18.08 \pm 0.04 $ & ATLAS \\ 
60517.19021 & g & AB & $ 18.33 \pm 0.10 $ & ZTF \\ 
60517.19021 & g & AB & $ 18.29 \pm 0.06 $ & ZTF \\ 
60518.08816 & c & AB & $ 18.29 \pm 0.04 $ & ATLAS \\ 
60518.26193 & o & AB & $ 18.02 \pm 0.04 $ & ATLAS \\ 
60519.05224 & c & AB & $ 18.25 \pm 0.040 $ & ATLAS \\ 
60519.17758 & g & AB & $ 18.34 \pm 0.09 $ & ZTF \\ 
60520.11622 & B &  Vega & $ 18.33 \pm 0.16 $ & Swift \\ 
60520.26668 & o & AB & $ 18.17 \pm 0.05 $ & ATLAS \\ 
60524.75334 & o & AB & $ 18.22 \pm 0.06 $ & ATLAS \\ 
60525.10893 & B &  Vega & $ 18.49 \pm 0.20 $ & Swift \\ 
60528.73356 & B &  Vega & $ 18.63 \pm 0.03 $ & LasCumbres \\ 
60528.73637 & V &  Vega & $ 18.06 \pm 0.04 $ & LasCumbres \\ 
60528.73825 & g & AB & $ 18.56 \pm 0.04 $ & LasCumbres \\ 
60528.74105 & r & AB & $ 18.38 \pm 0.05 $ & LasCumbres \\ 
60528.74294 & i & AB & $ 18.14 \pm 0.07 $ & LasCumbres \\ 
60536.07695 & B &  Vega & $ 18.63 \pm 0.05 $ & LasCumbres \\ 
60536.07975 & V &  Vega & $ 18.07 \pm 0.04 $ & LasCumbres \\ 
60536.08160 & g & AB & $ 18.58 \pm 0.03 $ & LasCumbres \\ 
60536.08439 & r & AB & $ 18.33 \pm 0.03 $ & LasCumbres \\ 
60536.08626 & i & AB & $ 18.12 \pm 0.04 $ & LasCumbres \\ 
60542.11852 & B &  Vega & $ 18.61 \pm 0.1 $ & LasCumbres \\ 
60542.12159 & V &  Vega & $ 17.87 \pm 0.06 $ & LasCumbres \\ 
60547.78241 & B &  Vega & $ 18.49 \pm 0.09 $ & LasCumbres \\ 
60547.78544 & V &  Vega & $ 18.0 \pm 0.12 $ & LasCumbres \\ 
60547.78753 & g & AB & $ 18.32 \pm 0.04 $ & LasCumbres \\ 
60548.10462 & B &  Vega & $ 18.76 \pm 0.05 $ & LasCumbres \\ 
60548.10769 & V &  Vega & $ 18.11 \pm 0.03 $ & LasCumbres \\ 
60548.10983 & g & AB & $ 18.5 \pm 0.05 $ & LasCumbres \\ 
60548.11288 & r & AB & $ 18.51 \pm 0.04 $ & LasCumbres \\ 
60548.11502 & i & AB & $ 18.16 \pm 0.05 $ & LasCumbres \\ 
60554.02099 & B &  Vega & $ 18.65 \pm 0.03 $ & LasCumbres \\ 
60554.02400 & V &  Vega & $ 18.14 \pm 0.03 $ & LasCumbres \\ 
60554.02609 & g & AB & $ 18.57 \pm 0.03 $ & LasCumbres \\ 
60554.02909 & r & AB & $ 18.45 \pm 0.03 $ & LasCumbres \\ 
60554.03115 & i & AB & $ 18.18 \pm 0.05 $ & LasCumbres \\ 
60563.37217 & B &  Vega & $ 18.69 \pm 0.06 $ & LasCumbres \\ 
60563.37498 & V &  Vega & $ 18.19 \pm 0.05 $ & LasCumbres \\ 
60563.37688 & g & AB & $ 18.59 \pm 0.04 $ & LasCumbres \\ 
60563.37970 & r & AB & $ 18.59 \pm 0.03 $ & LasCumbres \\ 
60563.38159 & i & AB & $ 18.24 \pm 0.04 $ & LasCumbres \\ 

\hline

\caption{Photometry of SN\,2024ahr. Magnitudes are corrected for Galactic Extinction in the direction of SN\,2024ahr.}
\label{tab:photometrytable}
\end{longtable}

\bibliography{ref.bib}{}
\bibliographystyle{aasjournal}

\end{document}